\documentclass{ws-procs9x6}

\def\to{\rightarrow}

\def\PL{{Phys.\ Lett.\ }}

\def\PRD{{Phys.\ Rev.\ D} }
\def\PRL{{Phys.\ Rev.\ Lett.\ }}

\def\de{\delta}

\def\ve{\varepsilon}

\def\th{\theta}

\def\ka{\kappa}

\def\rh{\rho}

\def\si{\sigma}

\def\cl{{\mathcal L}}

\def\fr#1#2{{{#1} \over {#2}}}
\def\half{{\textstyle{1\over 2}}}

\def\frac#1#2{{\textstyle{{#1}\over {#2}}}}

\def\lsim{\mathrel{\rlap{\lower4pt\hbox{\hskip1pt$\sim$}}
    \raise1pt\hbox{$<$}}}
\def\gsim{\mathrel{\rlap{\lower4pt\hbox{\hskip1pt$\sim$}}
    \raise1pt\hbox{$>$}}}
\def\sqr#1#2{{\vcenter{\vbox{\hrule height.#2pt
         \hbox{\vrule width.#2pt height#1pt \kern#1pt
         \vrule width.#2pt}
         \hrule height.#2pt}}}}

\def\prt{\partial}
\def\lrpartial{\raise 1pt\hbox{$\stackrel\leftrightarrow\partial$}}

\def\etal{{\it et al.}}

\newcommand{\beq}{\begin{equation}}
\newcommand{\eeq}{\end{equation}}
\newcommand{\bea}{\begin{eqnarray}}
\newcommand{\eea}{\end{eqnarray}}
\newcommand{\rf}[1]{(\ref{#1})}

\begin{document}

\title{Cosmology and spacetime symmetries}

\author{RALF LEHNERT}

\address{Department of Physics and Astronomy\\
Vanderbilt University,
Nashville, Tennessee, 37235\\ 
E-mail: ralf.lehnert@vanderbilt.edu}

\maketitle

\abstracts{
Cosmological models often contain scalar fields,
which can acquire global nonzero expectation values
that change with the comoving time.
Among the possible consequences of these scalar-field backgrounds,
an accelerated cosmological expansion, 
varying couplings,
and spacetime-symmetry violations
have recently received considerable attention.
This talk studies
the interplay of these three key signatures
of cosmologically varying scalars
within a supergravity framework.
}

\section{Introduction}

Sizable efforts are currently directed towards the search
for a more fundamental theory
that must resolve a variety of theoretical issues
left unaddressed in present-day physics.
Many approaches along these lines
require novel scalars as a key ingredient.
Moreover,
scalar fields are often invoked 
in cosmological models\cite{quint,kessence,infl}
to explain certain phenomenological questions,
such as the observed late-time accelerated expansion of the Universe,
the horizon and the flatness problem,
or the claimed variation of the fine-structure parameter.
The search for observational consequences of such scalars 
can therefore yield valuable insight into new physics.

Scalar fields can acquire global expectation values
that vary with the expansion of the Universe.
Measurable effects of variations in the scalar background are typically feeble
because of the cosmological time scales involved.
This suggests ultrahigh-precision studies
as a promising tool for experimental investigations.
The remaining task is
to identify suitable tests.

Varying scalars determine a spacetime-dependent background
that violates translation invariance.
This effect could be measurable 
because symmetries are typically amenable to high-precision tests.
Moreover,
translation-invariance breakdown
is typically associated with Lorentz violation\cite{klp03,blpr}
since translations, rotations, and boosts
are intertwined
in the Poincar\'e group.
This perhaps less appreciated result
opens a door for alternative experimental investigations
of cosmologically varying scalars
and is the primary focus of this talk.

Lorentz and CPT breakdown has also been suggested in other contexts 
as a candidate Planck-scale signature
including
strings,\cite{kps}
spacetime foam,\cite{ell98,suv}
nontrivial spacetime topology,\cite{klink}
loop gravity,\cite{amu}
and noncommutative geometry.\cite{chklo}
The emergent low-energy effects
are described by the Standard-Model Extension (SME),\cite{sme}
which
has provided the basis 
for studies
of Lorentz and CPT breaking
with 
mesons,\cite{hadronexpt,kpo,hadronth,ak}
baryons,\cite{ccexpt,spaceexpt,cane}
electrons,\cite{eexpt,eexpt2,eexpt3}
photons,\cite{photon}
muons,\cite{muons}
neutrinos,\cite{sme,neutrinos}
and the Higgs.\cite{higgs}

%Section \ref{arg} argues from a general viewpoint
%that varying scalars are typically associated
%with Lorentz and possibly CPT violation.
%In Sec.\ \ref{models},
%we consider a supergravity cosmology 
%leading to a fine-structure parameter $\alpha$
%and an electromagnetic $\theta$ angle
%that vary with time.
%Lorentz and CPT violation in the Maxwell and scalar sectors
%is discussed in Sec.\ \ref{lv}.

\section{Connection between translation and Lorentz symmetry}
\label{arg}

Consider the angular-momentum tensor $J^{\mu\nu}=\int d^3x \;\big(\th^{0\mu}x^{\nu}-\th^{0\nu}x^{\mu}\big)$,
which generates boosts and rotations.
Its construction 
involves the energy--momentum tensor $\th^{\mu\nu}$,
which is no longer conserved 
when spacetime-translation symmetry is broken.
Consequently,
$J^{\mu\nu}$
will typically depend on time,
so that the usual time-independent 
boost and rotation generators cease to exist.
As a result,
Lorentz and CPT invariance
is no longer assured.

Lorentz violation through varying scalars 
can also be understood in a more intuitive way.
A varying scalar is always associated 
with a nonzero 4-gradient.
This background gradient
determines a preferred direction on scales 
comparable to the variation:
Consider, 
for instance, 
a particle species
that is coupled to such a gradient.
The propagation features of these particles
might now depend upon whether the motion is perpendicular or parallel
to the background gradient.
This implies physically inequivalent directions,
and thus the violation of rotation invariance.
Since rotations are contained in the Lorentz group,
Lorentz symmetry must be broken.

We finally establish the effect at the Lagrangian level.
Consider
a model with a varying coupling $\xi(x)$
and two scalars $\phi$ and $\Phi$.
Suppose
that the Lagrangian  
contains a term 
$\xi(x)\,\partial^{\mu}\phi\,\partial_{\mu}\Phi$.
We next integrate the action
by parts with respect to the partial derivative acting on $\phi$.
This produces an equivalent Lagrangian 
$\mathcal{L}'\supset -K^{\mu}\phi\,\partial_{\mu}\Phi$.
Here $K^{\mu}\equiv\partial^{\mu}\xi$
is a nondynamical background 4-vector
violating Lorentz invariance.
Note
that for cosmological variations of $\xi$ 
we have $K^{\mu}={\rm const.}$ on small scales.

\section{Toy supergravity cosmology}
\label{models}

We now illustrate the result from Sec.\ \ref{arg}
with a toy model. 
This model is motivated by pure $N=4$ supergravity
in four spacetime dimensions.
Although unrealistic in detail,
it is a limit of $N=1$ supergravity 
in eleven dimensions,
which is contained in M-theory.
One thus expects
that our model
can illuminate generic aspects
of a candidate fundamental theory.

Suppose 
that only one of the model's graviphotons, 
$F^{\mu\nu}$, is excited.
Then, the bosonic part of pure $N=4$ supergravity
is given by \cite{cj,klp03}
\bea
\ka \cl_{\rm sg}
&=&
-\frac 1 2 \sqrt{g} R
+\sqrt{g} ({\prt_\mu A\prt^\mu A + \prt_\mu B\prt^\mu B})/{4B^2}
\nonumber\\
&&
\qquad\!
-\frac 1 4 \ka \sqrt{g} M F_{\mu\nu} F^{\mu\nu}
-\frac 1 4 \ka \sqrt{g} N F_{\mu\nu} \tilde{F}^{\mu\nu}\; ,
\label{lag2}
\eea
where $M$ and $N$ are known functions of the scalars $A$ and $B$.
%\beq
%M =\fr
%{B (A^2 + B^2 + 1)}
%{(1+A^2 + B^2)^2 - 4 A^2}\; ,
%\quad
%N = \fr
%{A (A^2 + B^2 - 1)}
%{(1+A^2 + B^2)^2 - 4 A^2}\; .
%\label{N}
%\eeq
The dual field-strength tensor 
is $\tilde{F}^{\mu\nu}=\ve^{\mu\nu\rh\si}F_{\rh\si}/2$
and $g=-\det (g_{\mu\nu})$,
as usual.
Note
that we can rescale
$F^{\mu\nu}\to F^{\mu\nu}/\sqrt{\ka}$
removing the explicit appearance 
of the gravitational coupling $\ka$
from the equations of motion.

For phenomenological reasons,
we also include
$\delta\cl=-\half \sqrt{g} (m_A^2 A^2+ m_B^2 B^2)$
into $\cl_{\rm sg}$.\cite{blpr}
We further represent the model's fermions\cite{cj}
by the energy--momentum tensor of dust
$T_{\mu\nu} = \rh\,u_\mu u_\nu$
describing, e.g., galaxies.
Here,
$u^\mu$ is a unit timelike vector
and $\rh$ is the fermionic energy density.

The usual assumption of an isotropic homogeneous
flat Friedmann--Robertson--Walker Universe
implies that $F^{\mu\nu}=0$ on large scales.
Our cosmology is then governed
by the Einstein equations
and the equations of motion for the scalars $A$ and $B$.
At tree level,
the fermionic matter is not coupled to the scalars,
so we can take $T_{\mu\nu}$
as covariantly conserved separately.
Searching for solutions with this input
yields a nontrivial dependence of $A$ and $B$ 
on the comoving time $t$.

\section{Lorentz violation}
\label{lv}

Consider small localized excitations of $F_{\mu\nu}$
in the scalar background $A_{\rm b}$ and $B_{\rm b}$
from Sec.\ \ref{models}.
Since experiments are usually confined
to a small spacetime region,
it is appropriate to work in local inertial coordinates.
The effective Lagrangian $\cl_{\rm cosm}$ for such situations
follows from Eq.\ \rf{lag2} and is
\beq
\cl_{\rm cosm}=
-\frac 1 4 M_{\rm b}F_{\mu\nu} F^{\mu\nu}
-\frac 1 4 N_{\rm b} F_{\mu\nu} \tilde{F}^{\mu\nu}\; ,
\label{efflag}
\eeq
where $A_{\rm b}$ and $B_{\rm b}$ 
imply the time dependence of $M_{\rm b}$ and $N_{\rm b}$.
Comparison with the usual Maxwell Lagrangian $\cl_{\rm em} =
-\fr{1}{4 e^2} F_{\mu\nu}F^{\mu\nu}
- \fr{\th}{16\pi^2} F_{\mu\nu} \tilde{F}^{\mu\nu}$
shows that $e^2 \equiv 1/M_{\rm b}$ and  $\th \equiv 4\pi^2 N_{\rm b}$.
Thus, 
$e$ and $\th$
acquire time dependencies,
as they are determined by the varying background
$A_{\rm b}$ and $B_{\rm b}$.

The breakdown of Lorentz symmetry
in our effective Lagrangian \rf{efflag}
is best exhibited
by the resulting modified Maxwell equations:
\beq
\fr{1}{e^2}\partial^{\mu}F_{\mu\nu}
-\fr{2}{e^3}(\partial^{\mu}e)F_{\mu\nu}
+\fr{1}{4\pi^2}(\partial^{\mu}\th)\tilde{F}_{\mu\nu}=0\; .
\label{Feom}
\eeq
In our toy cosmology,
the gradients of $e$ and $\th$
are nonzero,
approximately constant locally,
and act as a nondynamical external background.
Thus, the gradients select a preferred direction
in local inertial frames
violating Lorentz invariance.
The term containing $\partial^{\mu}\th$
can be identified 
with the $k_{AF}$ operator in the minimal SME.
This term has recently received a lot of attention.\cite{mcs}
For instance, 
it can lead to vacuum \v{C}erenkov radiation.\cite{cer}

Next,
we consider small localized excitations $\de A$ and $\de B$
of the scalar background in local inertial coordinates at the point $x_0$.
With the ansatz
$A(x)= A_{\rm b}(x)+\de A(x)$
and $B(x)= B_{\rm b}(x)+\de B(x)$
in the equations of motion for $A$ and $B$,
we find the linearized equations
\bea
{}\!\!\!0 &=& \big[\Box -2B^{\mu}\prt_{\mu}
+2m_A^2B_{\rm b}^2\big]\de A
-\big[2A^{\mu}\prt_{\mu}
-2A^{\mu}B_{\mu}-4m_A^2A_{\rm b}B_{\rm b}\big]\de B\; ,\nonumber\\
0 &=& \big[2A^{\mu}\prt_{\mu}\big]\de A
+\big[\Box  -2B^{\mu}\prt_{\mu}
+6m_B^2B_{\rm b}^2-A^{\mu}A_{\mu}+B^{\mu}B_{\mu}\big]\de B\; ,
\label{ablineq}
\eea
where
$A_{\rm b}$, $B_{\rm b}$,
$A^{\mu}\equiv B_{\rm b}^{-1}\prt^{\mu}A_{\rm b}$,
and $B^{\mu}\equiv B_{\rm b}^{-1}\prt^{\mu}B_{\rm b}$
are evaluated at $x_0$.

Equation \rf{ablineq} 
governs the propagation
of $\de A$ and $\de B$
in our cosmological background.
Note that
$A^{\mu}$ and $B^{\mu}$
are external nondynamical vectors
violating Lorentz symmetry.
This result also applies
to quantum theory:
the excitations $\de A$ and $\de B$
would be seen as the effective particles
corresponding to $A$ and $B$,
so that such particles would break Lorentz invariance.

\section*{Acknowledgments}
This work was funded 
by the Funda\c{c}\~ao Portuguesa para o Desenvolvimento.

\end{document}